\documentclass[journal,twoside]{IEEEtran}
\usepackage{graphicx} 
\usepackage{cite}
\usepackage{comment}
\usepackage{balance}
\usepackage{acronym}
\usepackage{amsmath}
\usepackage{tabularx}
\usepackage[dvipsnames]{xcolor}
\usepackage[hang]{subfigure}
\usepackage{booktabs}
\usepackage{tikz}
\usetikzlibrary{positioning}
\usetikzlibrary{shapes.multipart}
\usepackage{amsfonts}
\usepackage{enumitem,bbm}
\usepackage{wrapfig,hyperref}
\usepackage{soul}
\newcommand{\noteb}[1]{\textcolor{blue}{#1}}

\def\x{\mathbf{x}(t,k)}

\def\sq{s_q(t,k)}
\def\sd{s_d(t,k)}
\def\si{s_i(t,k)}
\def\hq{\mathbf{h}_q(k)}

\def\hn{\mathbf{h}_n(k)}

\acrodef{STFT}{short-time Fourier transform}
\acrodef{ISTFT}{inverse short-time Fourier transform}
\acrodef{BSS}{blind source separation}
\acrodef{DOA}{direction of arrival}
\acrodef{SC}{single channel}
\acrodef{DPRNN}{dual-path recurrent neural network}
\acrodef{DNN}{deep neural network}
\acrodef{TF}{time-frequency}
\acrodef{RTF}{relative transfer function}
\acrodef{ATF}{acoustic transfer function}
\acrodef{SI-SDR}{scale-invariant signal-to-distortion ratio}
\acrodef{MSE}{mean square error}
\acrodef{DFT}{discrete Fourier transform }
\acrodef{SIR}{signal-to-interference ratio}
\acrodef{OVA}{overlap-and-add}
\acrodef{SDR}{signal-to-distortion ratio}
\acrodef{BLSTM}{Bidirectional Long Short-Term Memory}
\acrodef{SOTA}{state-of-the-art}
\acrodef{RI}{Real-Imaginary}
\acrodef{RIR}{room impulse response}
\acrodef{SNR}{signal-to-noise ratio}
\acrodef{RNN}{Recurrent Neural Networks}
\acrodef{SE}{speaker extraction}
\acrodef{MVDR}{minimum variance distortion beamformer}
\acrodef{CASA}{computational simultaneous grouping scene analysis}
\acrodef{FC}{fully connected}
\acrodef{E2E}{end-to-end}
\acrodef{SDR}{signal to distortion ration}
\acrodef{SIR}{signal-to-interference ratio}
\acrodef{STOI}{short-time objective intelligibility}
\acrodef{PESQ}{perceptual evaluation of speech quality}
\acrodef{CNN}{convolutional neural network}
\acrodef{DSB}{delay and sum beamformer}
\acrodef{LCMV}{linear constraint minimum variance}
\acrodef{GEVD}{generalized eigen-value decomposition}
\acrodef{EVD}{eigen-value decomposition}
\acrodef{IPD}{inter-channel phase difference}

\acrodef{TSE}{target speaker extraction}

\begin{document}

\title{End-to-End Multi-Microphone Speaker Extraction Using  Relative Transfer Functions }
\author{Aviad Eisenberg, Sharon Gannot, Shlomo E. Chazan}

\author{Aviad Eisenberg, Sharon Gannot,~\IEEEmembership{Fellow,~IEEE}, and Shlomo E. Chazan
\thanks{A. Eisenberg (\texttt{aviad.eisenberg@biu.ac.il}) and S. Chazan (\texttt{shlomi.chazan@biu.ac.il}) are with Bar-Ilan University and OriginAI, Israel; S. Gannot (\texttt{sharon.gannot@biu.ac.il}) is with Bar-Ilan University, Israel.} 
\thanks{The work was partially supported by a grant from the Audition Project, Data Science Program, Council of Higher Education, Israel.}
}

\markboth{IEEE Signal Processing Letters,~Vol.~x, No.~y, Dec.~2024}
{Eisenberg \textit{et al.}: {E2E Deep BeamFormer for Speaker Extraction}}

\maketitle

\begin{abstract}
This paper introduces a multi-microphone method for extracting a desired speaker from a mixture involving multiple speakers and directional noise in a reverberant environment. In this work, we propose leveraging the instantaneous \ac{RTF}, estimated from a reference utterance recorded in the same position as the desired source. 
The effectiveness of the \ac{RTF}-based spatial cue is compared with \ac{DOA}-based spatial cue and the conventional spectral embedding. 
Experimental results in challenging acoustic scenarios demonstrate that using spatial cues yields better performance than the spectral-based cue and that the instantaneous \ac{RTF} outperforms the DOA-based spatial cue.
\end{abstract}


\section{Introduction}

The extraction of a desired speaker from multi-microphone mixtures containing multiple simultaneous speakers is essential in numerous modern applications and devices, including virtual assistants, hearing aids, and smartphones.

In recent years, multi-microphone algorithms based on \acp{DNN} have emerged. In \cite{FaSNet, IFaSNet, markovic22_interspeech}, the network directly processes multi-microphone signals to separate speakers in the scene. However, the absence of a beamformer structure complicates the explainability of the spatial properties of these approaches.

In another category of multi-microphone audio processing, known beamforming criteria or the beamformer weights are estimated directly \cite{ADLMVDR, DeepWithMVDR, NICEbeamRLR, CausalUnet, eabnet2022, Walter2022, Walter2023, Cohen2024Explainable, schwartz2024multi}. Studies, such as \cite{Walter2023, Cohen2024Explainable}, demonstrate that the spatial properties of the filter-and-sum operation can be preserved.

When side information about the desired speaker is available, the speaker separation task is referred to as \ac{TSE}. This side information can take various forms, such as a ``voice signature'' obtained during an enrollment stage, spatial information like the \ac{DOA} of the desired speaker, or visual cues, such as lip movements. A comprehensive survey on speech extraction methods is available in \cite{Zmolikova2023Extraction}. 
In \cite{delcroix2018single, Xu2019Ext, delcroix2020improving, Eisenberg2022Extraction, eisenberg2023two}, voice signatures are used to extract the desired source from a mixture recorded by a single microphone. This is typically achieved by inferring a speaker embedding from an enrollment stage.
In addition to the speaker embedding, speaker extraction can also leverage spatial information, usually inferred from multi-microphone measurements. The spatial information can be incorporated through \ac{DOA} estimation and/or by steering a beamformer \cite{vzmolikova2017learning,zorilua2021investigation,han2021multi,ge2022spex,elminshawi2023beamformer,gu2020multi,xu2020neural,xu2021generalized,tesch2023spatially,tesch2022tasl,teschtasl2024}. 
The works in \cite{teschtasl2024,tesch2022tasl} specifically analyze the spatial properties of the multi-microphone extraction method.




\ac{TSE} algorithms that employ ``voice signature'' for enrollment do not fully exploit spatial information, albeit processing multichannel data. Integrating the target speaker’s \ac{DOA} as a spatial cue may enhance performance. However, it is well established that \ac{RTF}-based beamformers typically outperform \ac{DOA}-based beamformers in reverberant environments \cite{Gannot2017}.

To address these limitations, we propose utilizing the instantaneous \ac{RTF} features of the desired speaker. Furthermore, we conduct an extensive comparison between the proposed \ac{RTF}-based method and alternative approaches that use spectral and \ac{DOA} enrollment features, along with a comparison to the \ac{MVDR} beamformer. Our results demonstrate that the proposed method consistently outperforms all other approaches.

\section{Problem Formulation}
A scenario comprising $Q$ concurrently active speakers, captured by $J$ microphones in a reverberant and noisy environment, is addressed. The problem is formulated in the \ac{STFT} domain, with $k \in \{0,\ldots, K-1\}$ and $t \in \{0,\ldots, T-1\}$ representing the frequency index and time-frame index, respectively, with $T$ and $K$ the total number of time-frames and frequency bands, respectively. 

The observed signal, as received by the microphone array, can be modeled as:
\begin{equation} 
\x = \sum_{q=1}^{Q} \hq \cdot \sq + n(t,k) \cdot \hn + \mathbf{v}(t,k),
\end{equation}
where $\sq$ denotes the clean, anechoic speech signal of the $q$-th speaker, $\hq$ is a $J \times 1$ vector comprising the time-invariant \acp{ATF} relating the $q$-th source and the microphone array, $n(t,k)$ is a directional noise source, $\hn$ is a $J \times 1$ vector comprising the \acp{ATF} relating the noise and the microphone array, and $\mathbf{v}(t,k)$ represents the sensor noise.

We focus on the scenario where only two concurrent speakers are present, namely $Q=2$, referred to as the desired speaker $\sd$ and the interference speaker $\si$.
Define the reverberant speech sources as received by the microphones as $\tilde{\mathbf{s}}_q(t,k) = s_q(t,k)  \mathbf{h}_q(k)$.

Denote $s_d^{\text{ref}}(t,k)$, the enrollment signal for $\sd$ and the respective reverberant signal as received by the microphone array as $\tilde{\mathbf{s}}_d^{\text{ref}}(t,k)$. 
Given the mixed signal $\x$ and the enrollment, we aim to extract a reverberant replica of the desired speaker, $\tilde{\mathbf{s}}_d(t,k)$.
We stress that the enrollment signal is uttered from the exact position of the desired source to provide the required spatial cue.

\begin{figure*}[htbp]
\centering
\tikzset{every picture/.style={line width=0.75pt}} 

\begin{tikzpicture}[x=0.75pt,y=0.75pt,yscale=-1,xscale=1]

\draw (44.64,245.51) node  {\includegraphics[width=38.12pt,height=35.67pt]{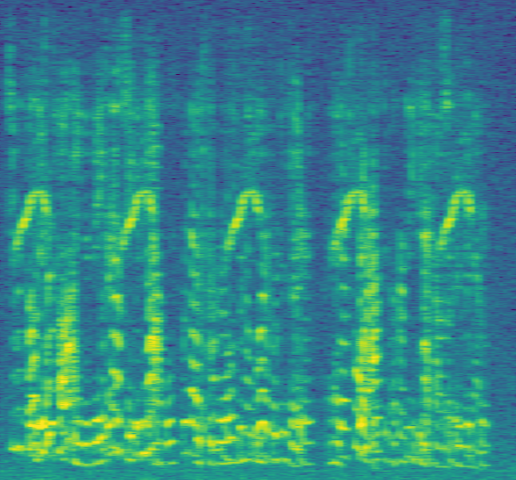}};
\draw   (19.23,221.73) -- (70.06,221.73) -- (70.06,269.29) -- (19.23,269.29) -- cycle ;

\draw (47.74,248.59) node  {\includegraphics[width=38.12pt,height=35.67pt]{mix.png}};
\draw   (22.33,224.81) -- (73.15,224.81) -- (73.15,272.37) -- (22.33,272.37) -- cycle ;

\draw (50.33,251.84) node  {\includegraphics[width=38.12pt,height=35.67pt]{mix.png}};
\draw   (24.92,228.05) -- (75.74,228.05) -- (75.74,275.62) -- (24.92,275.62) -- cycle ;

\draw (53.43,254.92) node  {\includegraphics[width=38.12pt,height=35.67pt]{mix.png}};
\draw   (28.01,231.14) -- (78.84,231.14) -- (78.84,278.7) -- (28.01,278.7) -- cycle ;

\draw   (156.1,369.84) -- (217.44,369.84) -- (217.44,386.25) -- (156.1,386.25) -- cycle ;
\draw   (156.1,386.25) -- (217.44,386.25) -- (217.44,402.67) -- (156.1,402.67) -- cycle ;
\draw   (156.1,419.08) -- (217.44,419.08) -- (217.44,435.5) -- (156.1,435.5) -- cycle ;
\draw  [fill={rgb, 255:red, 245; green, 166; blue, 35 }  ,fill opacity=1 ] (156.1,402.67) -- (217.44,402.67) -- (217.44,419.08) -- (156.1,419.08) -- cycle ;
\draw (210.08,428.82) node  {\includegraphics[width=8.07pt,height=8.67pt]{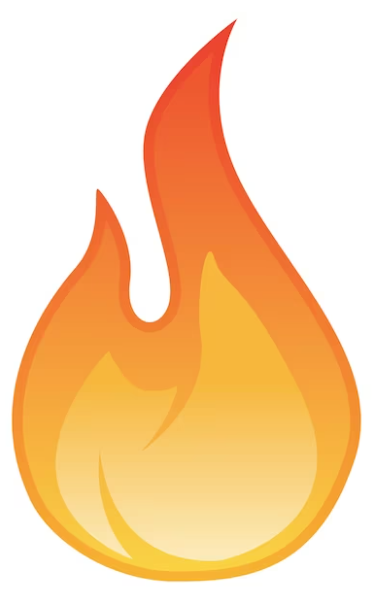}};

\draw    (60.98,402.67) -- (154.67,402.67) ;
\draw [shift={(156.67,402.67)}, rotate = 180] [color={rgb, 255:red, 0; green, 0; blue, 0 }  ][line width=0.75]    (10.93,-3.29) .. controls (6.95,-1.4) and (3.31,-0.3) .. (0,0) .. controls (3.31,0.3) and (6.95,1.4) .. (10.93,3.29)   ;
\draw    (95.1,395.29) -- (111.75,409.6) ;

\draw    (217.44,404.02) -- (234.26,404.02) -- (329.82,404.98) ;
\draw   (188.39,227.14) -- (248.75,244.9) -- (248.71,268.47) -- (188.31,286.06) -- cycle ;

\draw    (151.9,258.39) -- (186.06,258.39) ;
\draw [shift={(188.06,258.39)}, rotate = 180] [color={rgb, 255:red, 0; green, 0; blue, 0 }  ][line width=0.75]    (10.93,-3.29) .. controls (6.95,-1.4) and (3.31,-0.3) .. (0,0) .. controls (3.31,0.3) and (6.95,1.4) .. (10.93,3.29)   ;
\draw   (90.17,238.1) -- (151.51,238.1) -- (151.51,271.77) -- (90.17,271.77) -- cycle ;
\draw (46.17,331.03) node  {\includegraphics[width=38.12pt,height=35.67pt]{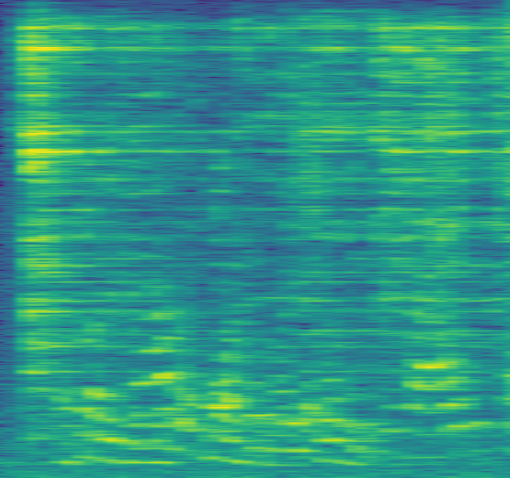}};
\draw   (20.75,307.24) -- (71.58,307.24) -- (71.58,354.81) -- (20.75,354.81) -- cycle ;

\draw   (83.6,321.36) .. controls (83.6,316.38) and (87.63,312.35) .. (92.61,312.35) -- (157.24,312.35) .. controls (162.22,312.35) and (166.25,316.38) .. (166.25,321.36) -- (166.25,348.38) .. controls (166.25,353.35) and (162.22,357.38) .. (157.24,357.38) -- (92.61,357.38) .. controls (87.63,357.38) and (83.6,353.35) .. (83.6,348.38) -- cycle ;
\draw (95.48,347.61) node  {\includegraphics[width=8.14pt,height=8.98pt]{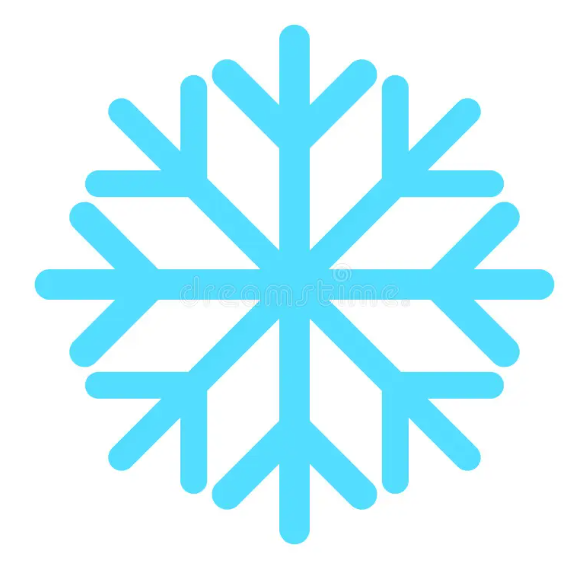}};

\draw   (177.7,318.11) -- (182.26,318.11) -- (182.26,346.31) -- (177.7,346.31) -- cycle ;
\draw   (191.39,318.11) -- (195.95,318.11) -- (195.95,346.31) -- (191.39,346.31) -- cycle ;
\draw   (186.82,318.11) -- (191.39,318.11) -- (191.39,346.31) -- (186.82,346.31) -- cycle ;
\draw   (182.26,318.11) -- (186.82,318.11) -- (186.82,346.31) -- (182.26,346.31) -- cycle ;
\draw   (195.95,318.11) -- (200.51,318.11) -- (200.51,346.31) -- (195.95,346.31) -- cycle ;

\draw   (403.89,243.52) -- (411.63,243.52) -- (411.63,283.07) -- (403.89,283.07) -- cycle ;
\draw   (248.21,164.74) -- (299.03,164.74) -- (299.03,212.3) -- (248.21,212.3) -- cycle ;
\draw (273.62,188.52) node  {\includegraphics[width=38.12pt,height=35.67pt]{mix.png}};

\draw   (250.79,167.5) -- (301.61,167.5) -- (301.61,215.06) -- (250.79,215.06) -- cycle ;
\draw (276.2,191.28) node  {\includegraphics[width=38.12pt,height=35.67pt]{mix.png}};

\draw   (254.23,170.74) -- (305.05,170.74) -- (305.05,218.31) -- (254.23,218.31) -- cycle ;
\draw (279.64,194.53) node  {\includegraphics[width=38.12pt,height=35.67pt]{mix.png}};

\draw   (256.81,173.5) -- (307.64,173.5) -- (307.64,221.06) -- (256.81,221.06) -- cycle ;
\draw (282.23,197.28) node  {\includegraphics[width=38.12pt,height=35.67pt]{mix.png}};

\draw   (318.45,168.85) -- (378.8,186.61) -- (378.77,210.18) -- (318.36,227.78) -- cycle ;

\draw   (394.3,184.29) -- (398.86,184.29) -- (398.86,212.48) -- (394.3,212.48) -- cycle ;
\draw   (407.99,184.29) -- (412.55,184.29) -- (412.55,212.48) -- (407.99,212.48) -- cycle ;
\draw   (403.43,184.29) -- (407.99,184.29) -- (407.99,212.48) -- (403.43,212.48) -- cycle ;
\draw   (398.86,184.29) -- (403.43,184.29) -- (403.43,212.48) -- (398.86,212.48) -- cycle ;
\draw   (412.55,184.29) -- (417.12,184.29) -- (417.12,212.48) -- (412.55,212.48) -- cycle ;

\draw    (407.22,242.63) -- (395.09,214.32) ;
\draw [shift={(394.3,212.48)}, rotate = 66.81] [color={rgb, 255:red, 0; green, 0; blue, 0 }  ][line width=0.75]    (10.93,-3.29) .. controls (6.95,-1.4) and (3.31,-0.3) .. (0,0) .. controls (3.31,0.3) and (6.95,1.4) .. (10.93,3.29)   ;
\draw    (407.22,242.63) -- (416.49,214.38) ;
\draw [shift={(417.12,212.48)}, rotate = 108.18] [color={rgb, 255:red, 0; green, 0; blue, 0 }  ][line width=0.75]    (10.93,-3.29) .. controls (6.95,-1.4) and (3.31,-0.3) .. (0,0) .. controls (3.31,0.3) and (6.95,1.4) .. (10.93,3.29)   ;
\draw    (407.22,242.63) -- (406.01,214.69) ;
\draw [shift={(405.92,212.69)}, rotate = 87.53] [color={rgb, 255:red, 0; green, 0; blue, 0 }  ][line width=0.75]    (10.93,-3.29) .. controls (6.95,-1.4) and (3.31,-0.3) .. (0,0) .. controls (3.31,0.3) and (6.95,1.4) .. (10.93,3.29)   ;

\draw   (493.25,227.85) -- (432.89,210.11) -- (432.92,186.54) -- (493.31,168.93) -- cycle ;
\draw (325.92,216.24) node  {\includegraphics[width=8.07pt,height=8.67pt]{hot.png}};
\draw (486.2,215.43) node  {\includegraphics[width=8.07pt,height=8.67pt]{hot.png}};
\draw   (501.92,172.69) -- (552.74,172.69) -- (552.74,220.25) -- (501.92,220.25) -- cycle ;
\draw (527.33,196.47) node  {\includegraphics[width=38.12pt,height=35.67pt]{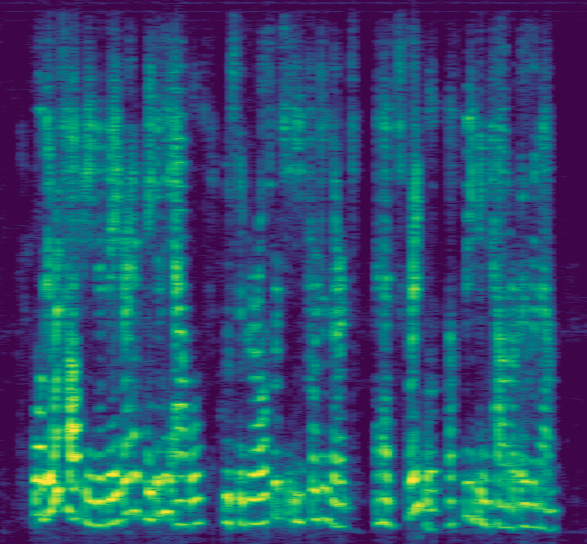}};

\draw   (257.36,241.36) -- (261.92,241.36) -- (261.92,269.55) -- (257.36,269.55) -- cycle ;
\draw   (271.05,241.36) -- (275.61,241.36) -- (275.61,269.55) -- (271.05,269.55) -- cycle ;
\draw   (266.49,241.36) -- (271.05,241.36) -- (271.05,269.55) -- (266.49,269.55) -- cycle ;
\draw   (261.92,241.36) -- (266.49,241.36) -- (266.49,269.55) -- (261.92,269.55) -- cycle ;
\draw   (275.61,241.36) -- (280.17,241.36) -- (280.17,269.55) -- (275.61,269.55) -- cycle ;

\draw    (407.82,283.09) -- (407.63,315.12) ;
\draw (196.02,274.11) node  {\includegraphics[width=8.07pt,height=8.67pt]{hot.png}};
\draw   (398.27,302.28) .. controls (400.87,302.27) and (402.99,304.36) .. (403,306.96) -- (403.07,321.08) .. controls (403.08,323.68) and (400.99,325.8) .. (398.39,325.81) -- (350.43,326.04) .. controls (350.43,326.04) and (350.43,326.04) .. (350.43,326.04) -- (350.32,302.5) .. controls (350.32,302.5) and (350.32,302.5) .. (350.32,302.5) -- cycle ;

\draw    (403.13,315.12) -- (407.63,315.12) ;
\draw    (329.82,404.98) -- (329.82,323.3) ;
\draw    (329.82,323.3) -- (349.06,323.79) ;
\draw [shift={(351.06,323.84)}, rotate = 181.46] [color={rgb, 255:red, 0; green, 0; blue, 0 }  ][line width=0.75]    (10.93,-3.29) .. controls (6.95,-1.4) and (3.31,-0.3) .. (0,0) .. controls (3.31,0.3) and (6.95,1.4) .. (10.93,3.29)   ;
\draw    (329.34,306.67) -- (347.91,307.04) ;
\draw [shift={(349.91,307.08)}, rotate = 181.13] [color={rgb, 255:red, 0; green, 0; blue, 0 }  ][line width=0.75]    (10.93,-3.29) .. controls (6.95,-1.4) and (3.31,-0.3) .. (0,0) .. controls (3.31,0.3) and (6.95,1.4) .. (10.93,3.29)   ;
\draw    (328.67,315.19) -- (347.91,315.68) ;
\draw [shift={(349.91,315.73)}, rotate = 181.46] [color={rgb, 255:red, 0; green, 0; blue, 0 }  ][line width=0.75]    (10.93,-3.29) .. controls (6.95,-1.4) and (3.31,-0.3) .. (0,0) .. controls (3.31,0.3) and (6.95,1.4) .. (10.93,3.29)   ;
\draw    (269.56,314.65) -- (328.67,315.19) ;
\draw    (226.22,334.66) -- (269.65,334.98) ;
\draw    (269.56,314.65) -- (269.65,334.98) ;
\draw    (329.72,257.99) -- (329.34,306.67) ;
\draw    (329.72,257.99) -- (307.91,258.17) ;

\draw (188.73,248.54) node [anchor=north west][inner sep=0.75pt]   [align=left] {Encoder};
\draw (286.54,249.93) node [anchor=north west][inner sep=0.75pt]    {$\sum $};
\draw (318.78,190.26) node [anchor=north west][inner sep=0.75pt]   [align=left] {Encoder};
\draw (286.67,221.55) node [anchor=north west][inner sep=0.75pt]   [align=left] {{\tiny [K,T,C]}};
\draw (57.61,279.42) node [anchor=north west][inner sep=0.75pt]   [align=left] {{\tiny [K,T,C]}};
\draw (55.66,355.69) node [anchor=north west][inner sep=0.75pt]   [align=left] {{\tiny [K,T]}};
\draw (536.39,221.14) node [anchor=north west][inner sep=0.75pt]   [align=left] {{\tiny [K,T]}};
\draw (94.71,244.83) node [anchor=north west][inner sep=0.75pt]    {$\widetilde{\boldsymbol{s}_{d}}\rightarrow \widehat{\boldsymbol{r}_{d}}$};
\draw (204.82,326.84) node [anchor=north west][inner sep=0.75pt]    {$\sum $};
\draw (351.64,303.82) node [anchor=north west][inner sep=0.75pt]  [rotate=-359.79] [align=left] {{\footnotesize arbitrator}};
\draw (40.81,395.88) node [anchor=north west][inner sep=0.75pt]    {$\theta $};
\draw (85.58,321.29) node [anchor=north west][inner sep=0.75pt]   [align=left] {SC Encoder};
\draw (434.1,190.39) node [anchor=north west][inner sep=0.75pt]   [align=left] {Decoder};

\end{tikzpicture}
\caption{The multichannel \ac{TSE} algorithm is described, comparing three alternative enrollment features: (1) the instantaneous \ac{RTF} extracted from clean speech utterances impinging on the array from the same position as the desired source, (2) a pre-trained single-channel model that generates a spectral embedding, and (3) the true \ac{DOA} of the desired source mapped to a learned representation. A separate model is trained for each enrollment feature.}
\label{fig:block_diagram}
\end{figure*}

\section{Proposed Model}
In this section, we present the proposed \ac{TSE} architecture, utilizing various types of enrollment features.
\subsection{Architecture}
\label{subsec:Architecture}

The proposed framework consists of a multi-channel encoder, a decoder, and an additional encoder for each enrollment feature. The multi-channel encoder processes the mixture signal, while the enrollment encoder extracts feature-specific embeddings. The decoder then utilizes both encoders to separate the target speaker.
Unlike \cite{eisenberg2023two}, Siamese encoders are not implemented in this study, as the input to both encoders has a different nature, as will be elaborated in the sequel.

The mixture encoder design employs multiple convolution layers, a two-dimensional batch normalization layer, and a `ReLU' activation function. Subsequently, the channel and frequency dimensions are merged, and a \ac{FC} layer is used to reduce dimensionality. A single self-attention layer is then applied. Finally, one of the alternative reference embedding vectors, described in the following paragraphs, is multiplied with each vector along the frame dimension of the mixture embedding on a frame-by-frame basis.

The decoder architecture consists of six self-attention layers and a \ac{FC} layer to restore the original dimension. To enable the use of skip connections as needed, transpose-convolution layers are employed to adapt to the convolution layers in the encoder. Finally, a self-attention layer is applied. 

A block diagram of the entire model is shown in Fig.~\ref{fig:block_diagram}.

\subsection{Alternative Enrollment Information}

In this study, we employed the \ac{RI} components of the \ac{STFT} as the features of the mixture signal. As for the reference features, we explored various alternatives. Two primary approaches emerged: utilizing prior spatial information or leveraging the speech attributes of our target speaker. We utilized either the instantaneous \ac{RTF} or the oracle \ac{DOA} in the former option. In the latter approach, we utilized the embedding of the \ac{SC} model, which does not involve spatial information.

\subsubsection{{RTF} Features}
While \ac{ATF} estimation is a blind problem, the \ac{RTF} estimation is a non-blind problem. The \ac{RTF}, proposed in \cite{gannot2001signal}, is widely used for beamforming and localization of sound sources.
The \ac{RTF} of the $q$-th source w.r.t.~microphone $m\in\{1,\ldots,J\}$ is defined as:
\begin{equation}
    \mathbf{r}_q(k) = \frac{1}{h_{m q}(k)} \hq. 
\end{equation}
The \ac{RTF} encodes the spatial information and will be utilized as an auxiliary feature for the proposed model.
%
%
%
In this study, assuming $\tilde{\mathbf{s}}_d^{\text{ref}}(t,k)$ is noiseless, we employ the instantaneous estimate of the \ac{RTF} \cite{chazan2019multi} as the ratio of the \ac{STFT} of the microphone signals:
\begin{equation}
\hat{\mathbf{r}}_d(t,k) = \frac{\tilde{\mathbf{s}}_d^{\text{ref}}(t,k)}{\tilde{s}_d^{\text{ref},m}(t,k)},
\end{equation}
where $\tilde{s}_d^{\text{ref},m}(t,k)$ represents the enrollment signal captured by the $m$-th microphone, an arbitrarily selected reference microphone. In our simulation study, we use the enrollment of the desired speaker, but in practice, any signal from the desired location can be utilized.
The encoder architecture for the instantaneous \ac{RTF} features parallels that of the mixture encoder described earlier. The encoder’s output, which is an embedding of the input features, is averaged across the frame dimension to yield a single representation vector, guiding the model toward the desired speaker.

\subsubsection{DOA Features}
As an alternative to the instantaneous \ac{RTF}, we provided the \ac{TSE} framework with the oracle \ac{DOA} of the desired speaker. It is important to note that, in real-world scenarios,  the \ac{DOA} must be estimated. In environments with high reverberation levels, accurate estimation becomes more challenging, which could impact the performance of the extraction model. Unlike the \ac{RI} features of the \ac{STFT} and \ac{RTF}, which share the same dimensions, the \ac{DOA} is an integer. Consequently, we modified the encoder architecture to allow it to learn the \ac{DOA} representation.

Using a lookup table, we introduce an embedding vector that is learned for each \ac{DOA}. Each \ac{DOA} is used to select a corresponding row in the table. This vector is then passed through a self-attention layer. Finally, the embedding vector is used in the bottleneck, as described in \ref{subsec:Architecture}.

\subsubsection{Spectral Features}

In addition to the previously discussed design options, we explored a scenario where spatial information about the desired speaker is unavailable, and the speaker's voice signature is provided instead. To achieve this, we employed the initial stage of the \ac{SC} model in \cite{eisenberg2023two} with the $m$-th microphone signal.
We then use the embedding vector derived from the enrollment signal as it was used with the \ac{DOA} and \ac{RTF} features. This approach shifted the model’s attention from the spatial information of the desired speaker to their speech characteristics.

\subsection{Objective function}

To train the proposed model for the extraction task, we used the time-domain \ac{SI-SDR} loss function, which is known to be effective in \ac{BSS} tasks. 
%

To further enhance the training process, we employed a technique where, for each training sample, we used the same mixture and swapped the desired and interference signals. The corresponding enrollment signal was then used for each of the extracted sources. The two losses were subsequently averaged, resulting in the following expression:
\begin{equation}
L_{\text{SI-SDR}} =   \frac{1}{Q}\sum_{q=1}^{Q}\text{SI-SDR}\left(\tilde{s}_q,\hat{\tilde{s}}_q \right) 
\label{eq:loss_sisdr}
\end{equation}

\section{Experimental Study}

\subsection{Datasets}
We evaluate the proposed approach on a simulated dataset. It consists of 40000 utterances for the training set, 5000 for
the validation set, and 300 for the test set. Each example is created by randomly choosing two utterances from the LibriSpeech database and convolving
the signals with four channel \ac{RIR}
simulated by the Image Method \cite{habets2006room}. 

The simulations were conducted in a room with dimensions uniformly distributed in $U[3,10]$~m and a reverberation time of $U[0.2,0.8]$~sec. Microphones were spaced 8~cm apart and placed at least 0.7~m away from the walls. Source positions were uniformly distributed in $U[0^\circ,180^\circ]$, with distances from the microphones in the range $U[1,4]$~m.

Furthermore, to enhance the model's spatial resolution, we introduced directional noise into the mixture with SNR drawn in the range $U[-5,20] $~dB. The noise source was extracted from the audiolabs' dataset\footnote{https://github.com/audiolabs/anechoic-noise} and convolved with  4-channels \ac{RIR} originating from the same room as the two primary speakers. The model’s task becomes more challenging when the noise exhibits directional characteristics because relying on spatial references to identify the desired speaker is complicated by the presence of another directive source. To simulate sensor noise, we introduced pink noise at an \ac{SNR} of 20~dB into the mixture, in addition to the directional noise.

For the mixture simulation, we randomly selected two signals from each speaker, designating one as the desired signal and the other as the enrollment. Both signals from each speaker underwent convolution with the same \ac{RIR}. This choice is based on the assumption that there are brief intervals where only the \noteb{noiseless} desired speaker is present, making it suitable for use as an enrollment signal. Both signals must bear the same spatial information to ensure spatial consistency between the desired source and the corresponding enrollment. The utterances of the desired and interference sources are summed together with the reverberant noise and sensor noise to form the mixture signal.




\subsection{Algorithm Settings}
The speech and noise signals were drawn from the database and downsampled to 8~[KHz]. The frame size of the \ac{STFT} is set to 256 samples with a $50\%$ overlap. Only the first 129 frequency bins are processed due to the symmetry of the \ac{DFT}.

For the training procedure, we employed the Adam optimizer~\cite{kingma2015adam} with a learning rate of 0.001 and a training batch size of 14. The weights are initialized randomly, and the signal lengths are varied randomly for each batch.

\subsection{Evaluation Measures}
To assess the effectiveness of the proposed algorithm, we employ two evaluation metrics:  \ac{SI-SDR} and  \ac{STOI} \cite{Taal2011STOI}. The former indicates the efficiency of speaker separation, while the latter indicates audio intelligibility.

\subsection{Compared Methods}
We compared our proposed methods with our prior work \cite{eisenberg2023two}, using only a single-channel input. Additionally, we compared the proposed methods with two variants of the \ac{MVDR} beamformer. The \ac{MVDR} beamformer aims to estimate the desired signal with minimal distortion while simultaneously minimizing noise. Denoting $\mathbf{g}_d(k)$ as the \ac{MVDR} beamformer at frequency $k$, the estimated signal is given by: 
\begin{equation}
    \hat{s}_d(k,t) = \mathbf{g}^{\text{H}}_d(k) \cdot \mathbf{x}(t,k),
\end{equation}
where $^\textrm{H}$ stands for the Hermitian operator. 
The \ac{MVDR} weights are given by:
\begin{equation}
    \label{eq:mvdr}
    \mathbf{g}_d(k) = \frac{ \mathbf{Q}^{-1}(k)\mathbf{r}^{\text{H}}_d(k)}{\mathbf{r}^{\text{H}}_d(k) \mathbf{Q}^{-1}(k) \mathbf{r}_d(k)},
\end{equation}
where $\mathbf{Q}(k)$ is the spatial covariance matrix of the noise and the interference.
The \ac{RTF}-based \ac{MVDR} aims at estimating the reverberant desired signal, as captured by a reference microphone. 
Given a directional noise and an additional (directional) speaker, we expect $\mathbf{Q}(k)$ to exhibit two main eigenvalues. Unlike the \ac{LCMV} beamformer, the \ac{MVDR} design does not include a constraint to place a null toward the interfering source. Instead, interference suppression is handled implicitly as part of the noise minimization.

We implemented two variants of the \ac{MVDR} beamformer. In the first variant, we assume the availability of a two-second interval containing only the desired speaker and noise signals, as well as an additional two-second segment containing only the noise signals (i.e., the interference is inactive). The \ac{RTF} of the desired speaker is then estimated using the covariance-whitening procedure \cite{Markovich-Golan2018Performance}. 
To estimate the spatial covariance matrix $\mathbf{Q}(k)$, we assume access to a segment containing interference plus the noises. The estimate is calculated as:
\begin{equation}
    \hat{\mathbf{Q}}(k) = \frac{1}{T}\sum_{t=0}^{T-1} \tilde{\mathbf{a}}(k,t) \cdot  \tilde{\mathbf{a}}^{\text{H}}(k,t)
\end{equation}
where $ \tilde{\mathbf{a}}(k,t) = \tilde{\mathbf{s}}_i(t,k) +  n(t,k) \cdot \hn + \mathbf{v}(t,k)$. This procedure constitutes the estimated \ac{MVDR} beamformer.

In the second variant, we assume that noiseless enrollments are available. Hence,  the spatial covariance matrix of the noise can be substituted with the identity matrix. This procedure constitutes the Oracle \ac{MVDR} beamformer.

\subsection{Results}
\begin{figure}[htbp]
\center
{\includegraphics[width=0.48\textwidth, height=0.5\textwidth]{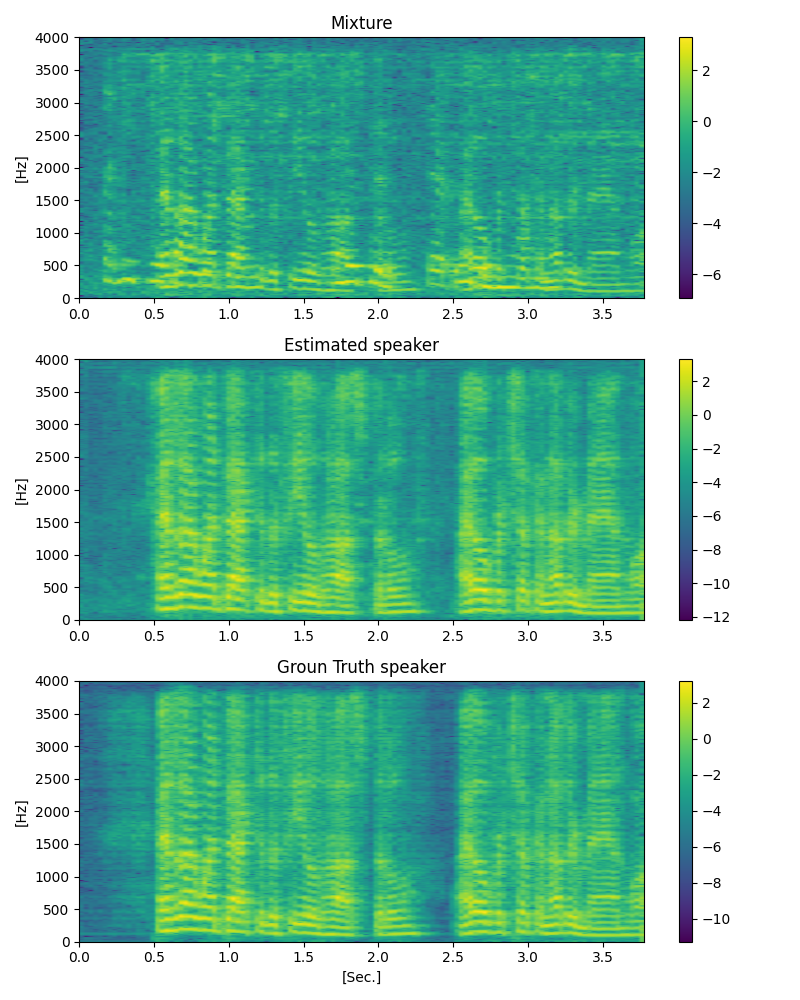}}
\caption{Spectrogram example of the Proposed-\ac{RTF} model compared with the noisy and ground-truth spectrogram.}
\label{fig:sonograms}
\end{figure}

The \ac{SI-SDR} and \ac{STOI} results for all methods evaluated on the dataset described above are shown in the middle part of Table~\ref{table:results}. All three proposed variants outperform the single-channel model and the estimated \ac{MVDR}. Among these variants, the algorithm using the instantaneous \ac{RTF} as a feature vector achieves the best performance. A spectrogram example of the proposed \ac{RTF}-based model is presented in Fig.~\ref{fig:sonograms}.

To further evaluate our model, we simulated a scenario where both speakers have the same \ac{DOA} but differ in their distances from the array. Specifically, we generated data where two speakers are aligned with the array’s center, while directional noise originates from a different direction. In this setup, a reference-based \ac{DOA} model would fail to distinguish the speakers due to their shared \ac{DOA}. Conversely, an \ac{RTF}-based approach can differentiate the sources, as it captures spatial information derived from multiple arrivals caused by room reverberation and signal propagation. We therefore assess the performance of both the Proposed-\ac{RTF} variant and the \ac{RTF}-based \ac{MVDR}.

The numerical results are in the bottom of Table~\ref{table:results}. The results show that the oracle \ac{MVDR} outperforms the proposed method in the intelligibility score, but our model excels when considering separation capabilities. The separation results are on par with those obtained for the case of different \acp{DOA}, which is quite remarkable.

\begin{table}[htbp]
\caption{Results for random locations (middle) and same \ac{DOA} (bottom).}
\begin{center}
\begin{tabular}{@{}lcccccc@{}}
\toprule
  Model & SI-SDR~[dB] $\uparrow$ &  STOI $\uparrow$     \\
\midrule
Unprocessed   & -2.6 & 0.54 \\

\midrule\midrule
Oracle \ac{MVDR}   & 9.7 &  0.85  \\
\midrule
Single channel   & 6.21 & 0.73 \\

Estimated \ac{MVDR} & 6.3 & 0.79 \\

TSE-RTF   & \textbf{9.2} & \textbf{0.81}  \\

TSE-DOA   & 8.4 & 0.8  \\

TSE-Spectral   & 8.18 &  0.8 \\

  \midrule\midrule

Oracle \ac{MVDR}   & 8.5 & \textbf{0.82} \\ 
TSE-\ac{RTF}   & \textbf{8.8} & 0.79 \\

\bottomrule
\end{tabular}
\end{center}
\label{table:results}
\end{table}


\section{Conclusions}
This study introduces a model for \acf{TSE} using a noiseless enrollment signal by utilizing the \ac{RTF} associated with the desired speaker. We evaluate the effectiveness of different features, specifically the known \ac{DOA} and spectral characteristics. Additionally, we present a comparison with the \ac{MVDR} beamformer. Our results demonstrate the benefits of leveraging the \ac{RTF}, even when both speakers originate from the same direction.

 \balance
\bibliographystyle{IEEEtran}
\bibliography{bib_short,main}

\end{document}